\documentclass[preprint2]{aastex}
\usepackage{amssymb,amsmath,amsfonts,graphicx,epsf}
\usepackage{color}
\begin{document}

%Title of paper
\title{Energy Dissipation in Magnetohydrodynamic Turbulence: Coherent Structures or ``Nanoflares''?}

\author{Vladimir Zhdankin$^1$, Stanislav Boldyrev$^1$, Jean Carlos Perez$^2$, and Steven M. Tobias$^3$} %\altaffilmark{1} 
\affil{$^1$ Department of Physics, University of Wisconsin-Madison \\
1150 University Avenue, Madison, Wisconsin 53706, USA}
\affil{$^2$ Space Science Center, University of New Hampshire, Durham, New Hampshire 03824, USA}
\affil{$^3$ Department of Applied Mathematics, University of Leeds, Leeds, LS2 9JT, UK}
%\email{aastex-help@aas.org}
%\date{today}

\date{\today}

\begin{abstract}
We investigate the intermittency of energy dissipation in magnetohydrodynamic (MHD) turbulence by identifying dissipative structures and measuring their characteristic scales. We find that the probability distribution of energy dissipation rates exhibits a power law tail with index very close to the critical value of $-2.0$, which indicates that structures of all intensities contribute equally to energy dissipation. We find that energy dissipation is uniformly spread among coherent structures with lengths and widths in the inertial range. At the same time, these structures have thicknesses deep within the dissipative regime. As the Reynolds number is increased, structures become thinner and more numerous, while the energy dissipation continues to occur mainly in large-scale coherent structures. This implies that in the limit of high Reynolds number, energy dissipation occurs in thin, tightly packed current sheets which nevertheless span a continuum of scales up to the system size, exhibiting features of both coherent structures and nanoflares previously conjectured as a coronal heating mechanism.
\end{abstract}


\section{Introduction} 

Turbulent astrophysical plasmas are typically associated with the complex morphology of magnetic field lines. In many important cases, the energy stored in the magnetic field can be comparable to or exceed the thermal energy of the plasma. Topological changes in the magnetic field structure, through the mechanism of magnetic reconnection, can then lead to the intense and intermittent release of magnetic energy into kinetic energy or thermal energy.  

Arguably the most famous example of this scenario is the Parker model of coronal heating \citep{parker1972,parker_1983,parker1988}. In this model, magnetic field lines anchored at both ends into the solar photosphere become increasingly tangled by photospheric motions, forming a braided structure. This is thought to cause a myriad of small-scale magnetic reconnection events, known as nanoflares, which may account for the observed heating of the solar corona. This process is generally modeled in the framework of magnetohydrodynamics (MHD) with line-tied boundary conditions and slow driving. Numerical simulations of this system show the production of current sheets along with power-law scaling relations \citep{dmitruk_gomez1999,einaudi_velli1999,rappazzo_etal2008, longcope_sudan1994, rappazzo_etal2010} and a power-law distribution of flare intensities \citep{buchlin_velli2007}. These numerical results are accompanied by analytic studies of stability \citep{ng_bhattacharjee1998, chieuh_zweibel1987, zweibel_li1987, huang_zweibel2009, delzanno_finn2008} and phenomenological models for the scaling of current sheet characteristics with resistivity \citep{cowley_etal1997, ng_bhattacharjee2008, uritsky2013}. 

One major goal of these studies is to reproduce and explain the observed distribution for solar flare intensities, which have a power law index near -1.8 during active times \citep{crosby_etal1993,aschwanden_etal_2000b} and possibly steeper than -2.0 for quiet times \citep{parnell_jupp2000}. Accurately measuring and explaining the index of this distribution is of great practical importance, since an index steeper than -2.0 is required for weak dissipative events, i.e. nanoflares, to dominate the overall heating of the corona \citep{hudson1991}. In this context, nanoflares are dissipative events with energy scales in the range of $10^{24} - 10^{27}$ ergs, much weaker than the typical observed solar flares with energies up to and exceeding $10^{30}$ ergs \citep{hudson1991}. The effect of the hypothesized nanoflare population is to give the background coronal emission a spiky character at small temporal and spatial scales \citep{parker1988}. It is often {\em assumed} that such nanoflares correspond to tiny, dissipation-scale current sheets.

The correlation between the intensity of the energy dissipation and the current sheet sizes is however nontrivial. In principle, relatively weak dissipation may occur throughout a long current sheet, while strong dissipation occupies a small, scattered fraction of the volume. The distribution of the energy dissipation over current sheets of various thicknesses, widths, and lengths is a difficult problem related to the intermittency of the plasma dynamics, caused by turbulence or other mechanisms such as self-organized criticality \citep{aschwanden_2012}.

More generally, the intermittency of energy dissipation and plasma heating is an essential ingredient for a broad range of other space and astrophysical systems. In high-energy astrophysical systems, inhomogeneous temperature profiles may arise when strong prompt radiation removes energy from localized dissipation sites more rapidly than it can be redistributed in the medium, affecting the thermodynamics of such systems \citep[e.g.,][]{dahlburg_etal2012}. Examples of such systems include quasars \citep[][]{goodman_uzdensky2008}, accretion disks and flows \citep[][]{pariev_etal2003,blaes2013}, and hot X-ray gas in galaxy clusters. In collisionless and weakly collisional plasmas, intermittency sets the distribution and coherence lengths of electric fields, contributing to nonthermal particle acceleration. This is relevant for systems such as radiatively-inefficient accretion flows, galaxy clusters, molecular clouds, and the solar wind. For example, magnetic discontinuities measured in the solar wind can potentially be explained as signatures of intermittent structures, which would then contribute to particle heating \citep{veltri1999, bruno2001,greco2010, zhdankin2012b}.

Recent increases in supercomputing power has enabled the testing of some fundamental ideas of intermittency in the Parker model. According to our discussion above, one of these questions is whether, in the limit of vanishing resistivity, magnetic energy is released in an increasing number of progressively weaker and smaller reconnection events (nanoflares), or instead remains concentrated in a few intense large-scale structures independently of the resistivity. This question has long been recognized to be of fundamental importance for the Parker model of the solar corona \citep{ng_bhattacharjee1998, ng_bhattacharjee2008, ng_etal2012, rappazzo_etal2013, at2012, at2013, lin2013}. In fact, it is an equally fundamental question for MHD turbulence in general \citep{einaudi_velli1999}.

In the present work, we investigate an analogous problem for the intermittency of energy dissipation in resistive MHD turbulence driven at large scales, rather than the Parker model. It has long been known that the nonlinear interactions in MHD turbulence lead to the formation of intense dissipative structures in the guise of current sheets \citep{biskamp2003, muller_biskamp2000, muller_etal2003}. However, the question of how their characteristics scale with Reynolds number has not been systematically treated in a quantitative manner. Therefore, we seek to determine whether energy dissipation in the high Reynolds number limit is dominated by weak and increasingly numerous small-scale structures or by a fixed number of large-scale coherent structures. Qualitatively, the question at hand is whether intermittency in the high Reynolds number limit is spiky and chaotic in space and bursty in time, or coherently self-organized in both space and time.

Our results are not applicable to all aspects of the solar corona dynamics due to the different boundary conditions and forcing mechanisms. They do, however, describe robust small-scale properties of critically-balanced MHD turbulence. To facilitate the discussion, we adopt a generalized definition of nanoflares based on the characteristic scales of structures relative to the dissipation scale. Specifically, we define a nanoflare to be a dissipative structure with scales comparable to the dissipation range, while a coherent structure is a dissipative structure with scales within (or larger than) the inertial range. Hence, when the Reynolds number is pushed to large values, nanoflares will become vanishingly small while coherent structures will remain macroscopic. The corresponding number of nanoflares must increase with Reynolds number. Although temporal scales are not explicitly referred to, nanoflares are implied to be short-lived while coherent structures are relatively long-lived. Under these definitions, a structure can be both a nanoflare and a coherent structure under some circumstances, e.g., in a highly anisotropic system.

In order to quantitatively address the posed question, we perform a series of numerical simulations of reduced MHD to investigate strong MHD turbulence with progressively increasing Reynolds number. We apply novel methods to identify and measure the characteristic scales of structures in the current density. We confirm that energy dissipation is dominated by thin current sheets with thicknesses that are deep within the dissipation range. We discover, however, that these structures have lengths and widths that span the inertial range. Furthermore, we find that the energy dissipation rate is distributed uniformly across structures of all intensities, lengths, and widths in the inertial range. As the Reynolds number is increased, the structures become thinner and more numerous, while their lengths and widths continue to occupy a continuum of inertial-range scales up to the system size. In this sense, structures in MHD turbulence exhibit features of both coherent structures and nanoflares.
\\

\section{Method}

We analyze numerical simulations of reduced MHD (RMHD) for incompressible strong MHD turbulence with a strong uniform guide field $\boldsymbol{B}_0 = B_0 \hat{\boldsymbol{z}}$. The ratio of guide field to the root-mean-square average of the fluctuating component is fixed at $B_0/b_\text{rms} \approx 5$. In this case, the field fluctuations are predominantly perpendicular to $\boldsymbol{B}_0$, and so the RMHD equations are valid,
\begin{eqnarray}
\left(\frac{\partial}{\partial t}\mp\boldsymbol{V}_A\cdot\nabla_\parallel\right)\boldsymbol{z}^\pm+\left(\boldsymbol{z}^\mp\cdot\nabla_\perp\right)\boldsymbol{z}^\pm \nonumber \\ = -\nabla_\perp P 
+\nu\nabla_\perp^2\boldsymbol{z}^\pm +\boldsymbol{f}_\perp^\pm
\label{rmhd-elsasser}
\end{eqnarray}  
and $\nabla_\perp \cdot \boldsymbol{z}^\pm = 0$, where $\boldsymbol{z}^\pm=\boldsymbol{v}\pm\boldsymbol{b}$ are the Els\"asser variables (which are strictly perpendicular to $\boldsymbol{B}_0$), $\boldsymbol{v}$ is the fluctuating plasma velocity, $\boldsymbol{b}$ is the fluctuating magnetic field (in units of the Alfv\'en velocity, $\boldsymbol{V}_A={\boldsymbol{B}}_0/\sqrt{4\pi\rho_0}$, where $\rho_0$ is plasma density), $P=(p/\rho_0+b^2/2)$, $p$ is the plasma pressure, $\nu$ is the fluid viscosity, assumed to be equal to the magnetic diffusivity $\eta$ for simplicity (i.e. the Prandtl number is $Pm = \nu/\eta = 1$), and $\boldsymbol{f}^\pm_\perp$ is the large-scale forcing. The current density in RMHD is a scalar field given by $j = (\nabla_\perp \times \boldsymbol{b})\cdot\hat{\boldsymbol{z}}$.

The RMHD equations (\ref{rmhd-elsasser}) are solved in a periodic, rectangular domain of size $L_\perp = 2\pi$ perpendicular to the guide field and size $L_\parallel = 6L_{\perp}$ parallel to the guide field (refer to \cite{perez2010,perez_etal2012} for details on simulations). The turbulence is driven at the largest scales by colliding Alfv\'en modes, excited by statistically independent random forces $\boldsymbol{f}^+$ and $\boldsymbol{f}^-$ in Fourier space at low wave-numbers $2\pi/L_{\perp} \leq k_{x,y} \leq 2 (2\pi/L_{\perp})$, $k_z = 2\pi/L_\|$. The Fourier coefficients of $\boldsymbol{f}^\pm$ in this range are Gaussian random numbers with amplitudes chosen so that $b_\text{rms}\sim v_\text{rms}\sim 1$. The forcing is solenoidal in the perpendicular plane and has no component along $\boldsymbol{B}_0$. The random values of the different Fourier components of the forces are refreshed independently on average about $10$ times per eddy turnover time. To perform the spatial discretization, a fully dealiased 3D pseudo-spectral algorithm is used. Reynolds number is given by $Re = b_{\rm rms} (L_\perp/2\pi) / \nu$. The analysis is performed for 15 snapshots (spaced at intervals of one eddy-turnover time) each for runs with $Re = 1000$, $Re = 1800$, and $Re = 3200$ on $1024^3$ lattices, and also for 9 snapshots with $Re = 9000$ on a $2048^3$ lattice. In addition, analysis was performed on lower-resolution $512^3$ simulations to establish numerical accuracy of the methods for the low Reynolds number cases.

\begin{figure}[t!]
 \centering
\includegraphics[width=\columnwidth]{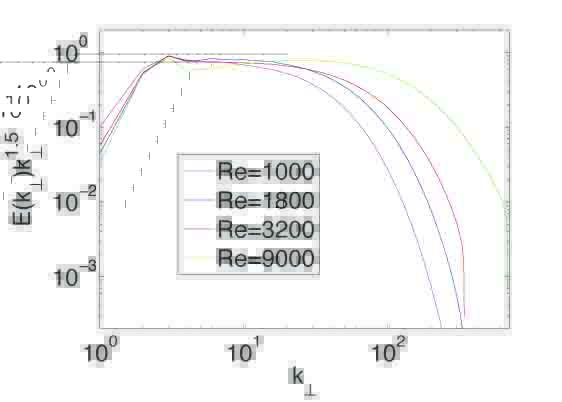}
 \includegraphics[width=\columnwidth]{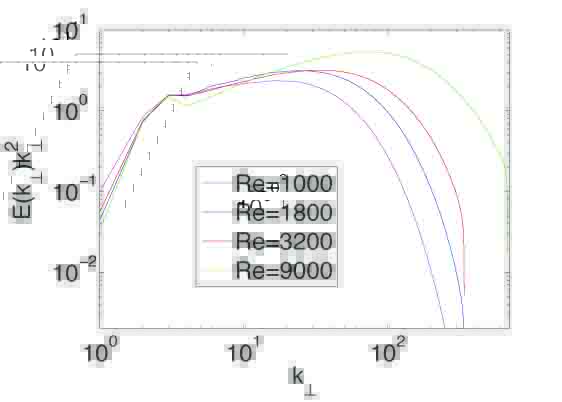}
\includegraphics[width=\columnwidth]{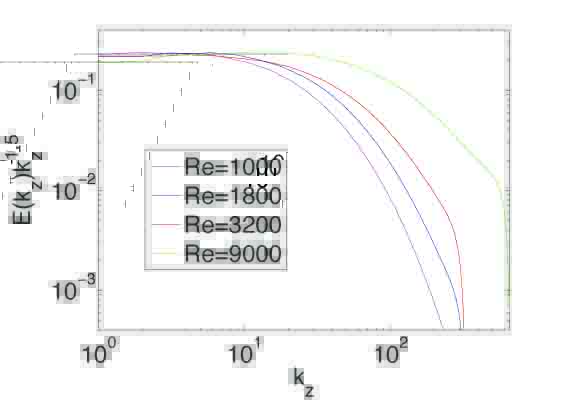}
 \caption{Top panel: Energy spectrum for perpendicular fluctuations in the magnetic field, compensated by $k_\perp^{3/2}$, for $Re = 1000$ (magenta), $Re = 1800$ (blue), $Re = 3200$ (red), and $Re = 9000$ (green). Center: Same spectrum compensated by $k_\perp^2$, representing the current density fluctuations. Bottom: Energy spectrum for magnetic field fluctuations in the $z$ direction, compensated by $k_z^{3/2}$. \label{fig:spec}}
 \end{figure}

For reference, in Fig.~\ref{fig:spec} we show the perpendicular magnetic energy spectrum averaged over the given snapshots, compensated by $k_\perp^{3/2}$. The magnetic energy spectrum clearly exhibits an inertial range which increases in size with Reynolds number. By compensating by an additional factor of $k_\perp^{1/2}$, as shown in the second panel of Fig.~\ref{fig:spec}, the energy spectrum for the current density is obtained, which peaks at wavenumbers beyond the inertial range. Hence, the energy spectrum requires the bulk of energy dissipation to occur in smaller and smaller scales as Reynolds number increases. We also show in Fig.~\ref{fig:spec} the magnetic energy spectrum in the $z$ direction, compensated by $k_z^{3/2}$, which better represents the perpendicular cascade rather than the parallel cascade, as noted in past studies e.g. \cite{maron_goldreich_2001}.

In order to study dissipative structures in a robust and quantitative manner, we apply the following algorithm. We set a threshold current density $j_\text{thr}$ and determine sets of spatially-connected points satisfying $|j| > j_\text{thr}$. Two points on the lattice are considered spatially-connected if one is contained in the other's 26 nearest neighbors. We then identify each of these non-intersecting point sets as a structure. Note that some structures with $|j| \approx j_\text{thr}$ will inevitably be under-resolved, but these represent a negligible fraction of energy dissipation and can be distinguished from resolved structures by their small scales. The energy dissipation rate of a given structure is given by ${\cal E} = \int dV \eta j^2$, where integration is performed across the points constituting the structure.

\begin{figure}[t!]
 \centering
 \includegraphics[width=\columnwidth]{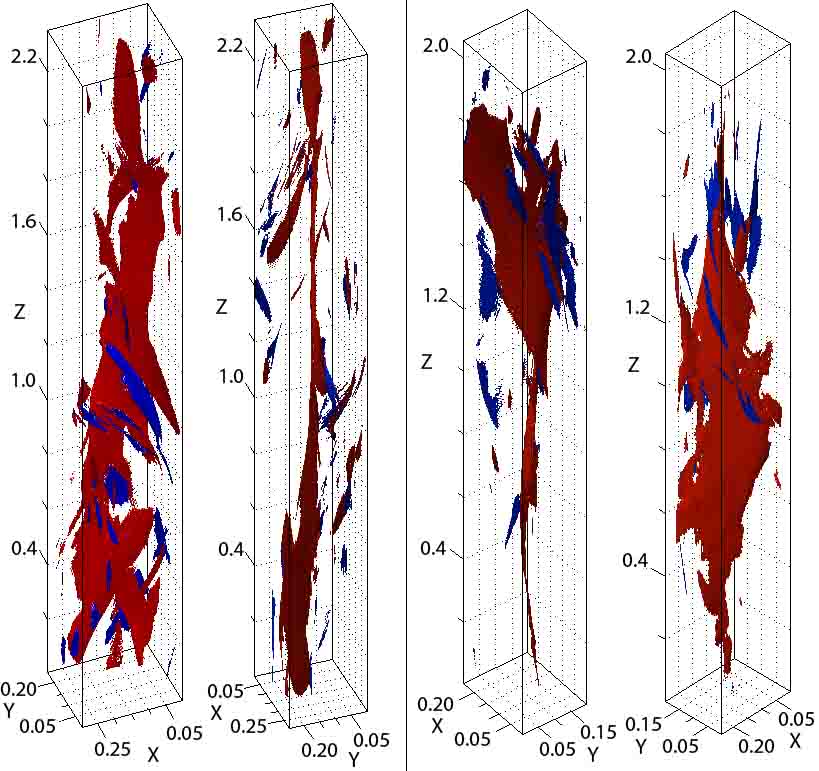}
 \caption{Samples of typical large current sheets in part of the simulation domain (in red), surrounded by several smaller structures (mostly in blue). The left panel shows two orientations of one structure, while the right panel shows two orientations of another separate structure. These samples are taken from the $Re = 1800$ case with a threshold of $j_\text{thr}/j_\text{rms} \approx 6.5$. \label{fig0}}
 \end{figure}

\begin{figure}[t!]
 \centering
 \includegraphics[width=0.75\columnwidth]{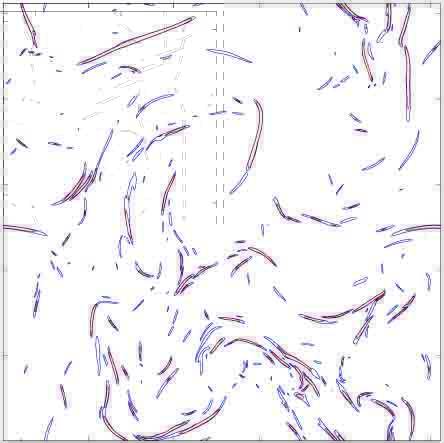}
\includegraphics[width=0.75\columnwidth]{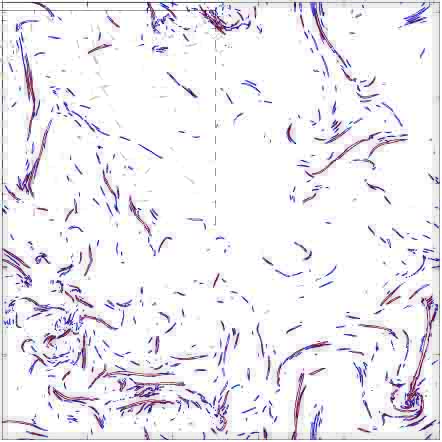}
\includegraphics[width=0.75\columnwidth]{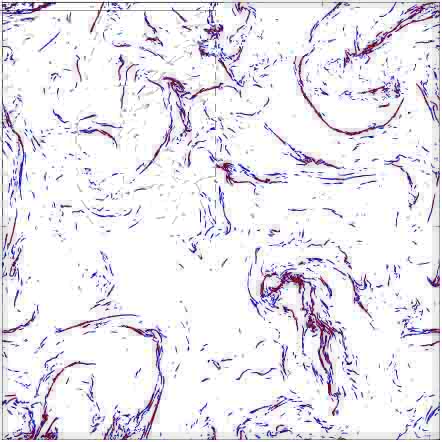}
 \caption{Contours of current density in an arbitrary plane perpendicular to the guide field. Contours are taken at $j_\text{thr}/j_\text{rms} = 2$ (blue) and $j_\text{thr}/j_\text{rms} = 3$ (red) for increasing Reynolds number (from top to bottom, $Re = 1000$, $3200$, and $9000$). \label{fig0b}}
 \end{figure}

Shown in Fig.~\ref{fig0} are two samples of large current sheets in part of the simulation domain, identified using the threshold procedure. Each structure is shown from two orientations, demonstrating the ribbon-like shape of the structures. Shown in Fig.~\ref{fig0b} are contours of current density in an arbitrary plane perpendicular to the guide field. The three panels show increasing Reynolds number, in the order $Re = 1000$, $Re = 3200$, and $Re = 9000$. These contour plots reveal that there is finer structure with more complex morphology when $Re$ increases.

Each structure is characterized by three characteristic scales: the length $L$, width $W$, and thickness $T$, with $L \ge W \ge T$. We apply two methods to measure the characteristic scales of each structure. The first method is based on the direct measurement of distance across the structure in three orthogonal directions, while the second method is based on the ratios of the Minkowski functionals \citep{kerscher2000}. We will refer to these as the Euclidean scales and the Minkowski scales, respectively. The Euclidean scales are intuitive local measurements of scale which may be misleading for irregular morphologies, whereas the Minkowski scales are mathematically rigorous measurements which are better applicable to complex morphologies, but may elude a straightforward physical interpretation.

We first describe the Euclidean method. For length $L_e$, we take the maximum distance between any two points in the structure. For width $W_e$, we consider the plane orthogonal to the length and coinciding with the point of peak current density. We then take the maximum distance between any two points of the structure in this plane to be the width. The direction for thickness $T_e$ is then set to be orthogonal to length and width. We take the thickness to be the distance across the structure in this direction through the point of peak current density. Since typical thicknesses may be comparable to the lattice spacing, we use a linear interpolation scheme to obtain finer measurements.

We now describe the Minkowski method, which has previously been applied to study the morphology of large-scale structures in the universe \citep{schmalzing_etal1999}, coherent structures in the kinematic dynamo \citep{wilkin_etal2007}, and vorticity filaments in hydrodynamic turbulence \citep{leung_etal2012}. By Hadwiger's theorem, the morphology of an object in $d$-dimensional space is completely characterized by the set of $d+1$ numbers known as the Minkowski functionals \citep{mecke2000}. In three-dimensional space, the first three Minkowski functionals are given by
\begin{eqnarray}
V_0 = V &=& \int dV \\
V_1 = \frac{A}{6} &=& \frac{1}{6} \int dS \\
V_2 = \frac{H}{3 \pi} &=& -\frac{1}{6 \pi} \int dS \nabla \cdot \hat{n}
\end{eqnarray}
where $V$ is volume, $A$ is surface area, and $H$ is the mean curvature on the surface (and $\hat{n} = \nabla j / |\nabla j|$ is the surface normal). Note that there also exists a fourth Minkowski functional $V_3 = \chi$, the Euler characteristic, but it will not be used here since it is dimensionless. Three quantities with the dimensions of length can be formed from ratios of these functionals,
\begin{eqnarray}
L_m &=& \frac{3 V_2}{4} \\
W_m &=& \frac{2 V_1}{\pi V_2} \\
T_m &=& \frac{V_0}{2 V_1}
\end{eqnarray}
where normalizations are chosen such that when applied to a sphere, all scales correspond to the radius. For simple convex objects, these three scales have the usual interpretation of length, width, and thickness.

To compute the Minkowski functionals on a lattice, we employ Crofton's formula, as described in \cite{schmalzing_buchert1997}. This method is based on counting the number of lattice points, lattice edges, lattice faces, and lattice cubes that constitute the structure. Accuracy of the Crofton method was established on low-resolution ($512^3$) simulations by comparing it with another numerical method, based on Koenderink invariants, also discussed in \cite{schmalzing_buchert1997}.

\section{Results}

\begin{figure}[t!]
 \includegraphics[width=\columnwidth]{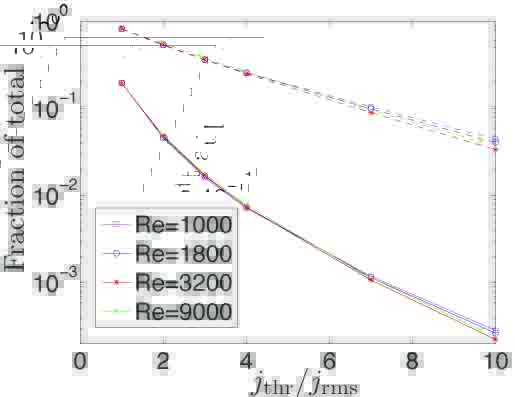}
 \caption{The fraction of total energy dissipation (dashed lines) and fraction of total volume (solid lines) accounted for by structures with current densities $|j| > j_\text{thr}$. The $x$-axis is the threshold relative to $j_\text{rms} = (E_\text{tot}/ \eta V_\text{tot})^{1/2}$, which evidently gives a universal result. The colors correspond to $Re = 1000$ (magenta), $Re = 1800$ (blue), $Re = 3200$ (red), and $Re = 9000$ (green).\label{fig1}}
 \end{figure}

We first discuss the dependence of our results on the threshold $j_\text{thr}$. Note that the rms current density diverges as $Re$ increases, since $j_\text{rms} = \sqrt{E_\text{tot}/\eta V_\text{tot}} \propto Re^{1/2}$, where system energy dissipation rate $E_\text{tot} = \int dV \eta j^2 \approx 1$ and system volume $V_\text{tot} =L_\perp^2 L_\parallel = 6(2\pi)^3$ are fixed. Therefore, we use the rescaled threshold $j_\text{thr}/j_\text{rms}$ to study the field in a universal manner.  As shown in Fig.~\ref{fig1}, the fraction of total energy dissipation accounted for by structures with $|j| > j_\text{thr}$ increases approximately as an exponential as the threshold decreases. This result is evidently universal in the variable $j_\text{thr}/j_\text{rms}$. The fraction of volume occupied is much smaller than the fraction of energy dissipated; for example, $40\%$ of energy is dissipated in approximately $2\%$ of the volume. In the following analysis, we choose $j_\text{thr}/j_\text{rms} \approx 3.75$ for all cases, giving a similar combined energy dissipation rate and volume occupied for structures independently of~$Re$. This threshold is chosen low enough to get a large sample of structures while being high enough to avoid many structures percolating through the domain. The results are similar for different thresholds as long as thresholds are several times larger than the rms; other statistical properties such as distributions of the scales and correlations between the scales also do not change significantly.

\begin{figure}[t!]
 \centering
\includegraphics[width=\columnwidth]{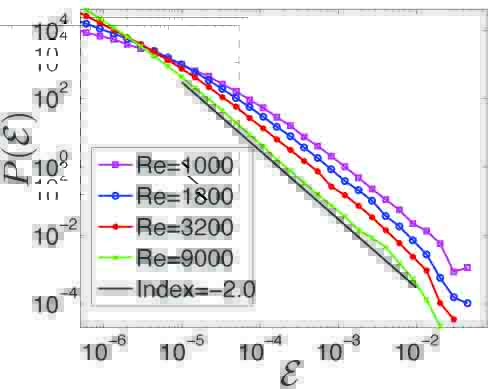}
\includegraphics[width=\columnwidth]{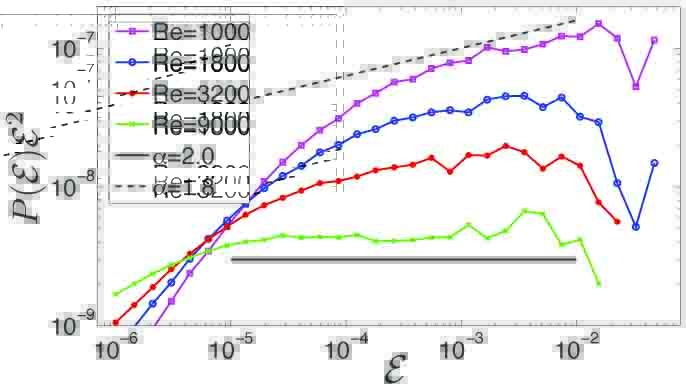}
\includegraphics[width=\columnwidth]{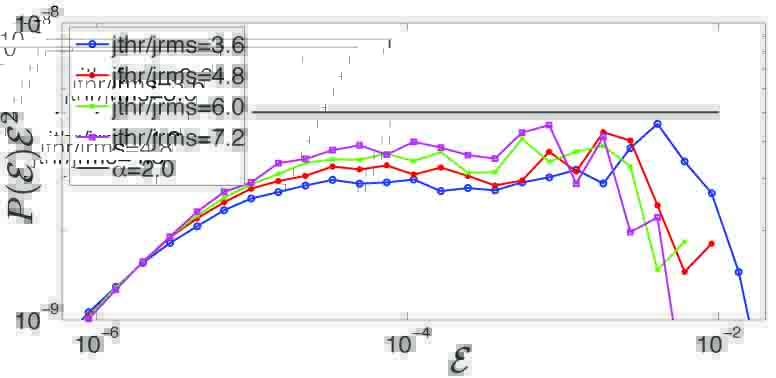}
 \caption{Top panel: the probability distribution $P({\cal E})$ for energy dissipation rate of structures, with colors corresponding to $Re=1000$ (magenta), $Re = 1800$ (blue), $Re = 3200$ (red), and $Re = 9000$ (green). The index for the power-law tail becomes increasingly close to the critical value of $-2$ as $Re$ increases. Middle panel: the same distribution compensated by ${\cal E}^2$, better showing the convergence with $Re$. Bottom panel: the compensated distribution for $Re = 9000$ with several different thresholds, $j_\text{thr}/j_\text{rms} = 3.6$ (blue), $4.8$ (red), $6.0$ (green), and $7.2$ (magenta). \label{figdist}}
 \end{figure}

We now consider the probability distribution for energy dissipation rates in the given population of structures. Let $P({\cal E}) d{\cal E}$ denote the number of structures with energy dissipation rate between ${\cal E}$ and ${\cal E} + d{\cal E}$, normalized to the total number. As shown in Fig.~\ref{figdist}, the distribution has a power-law tail $P({\cal E}) \sim {\cal E}^{-\alpha}$ with an index between $\alpha = 1.8$ and $\alpha = 2.0$ for all cases. From the compensated distribution $P({\cal E}) {\cal E}^2$, it is clear that with increasing $Re$, the distribution approaches the critical index $\alpha = 2.0$. This index is independent of the threshold, as demonstrated for the $Re = 9000$ case in the final panel of Fig.~\ref{figdist}. For the case with lower $Re$, the apparent index is closer to $-1.8$, which is consistent with several past studies \citep{uritsky2010, zhdankin_etal2013,buchlin_velli2007} and similar to the observed distribution of solar flare energies in the solar corona \citep{crosby_etal1993}.  A distribution with the critical index has an expected energy dissipation rate $\langle {\cal E} \rangle = \int d{\cal E} P({\cal E}) {\cal E}$ which is marginally divergent at both limits. Therefore, energy dissipation is distributed uniformly across structures of all intensities in this range, with no preference toward intense structures or weak structures \citep{hudson1991}. This attractive result was not revealed in previous studies of driven MHD turbulence, possibly because of low Reynolds number. In the regime of small ${\cal E}$, $P({\cal E})$ becomes shallower with no evident universal behavior. The structures in this regime may be a combination of structures near the threshold and structures completely within the dissipation range.

\begin{figure*}[t!]
 \includegraphics[width=2\columnwidth]{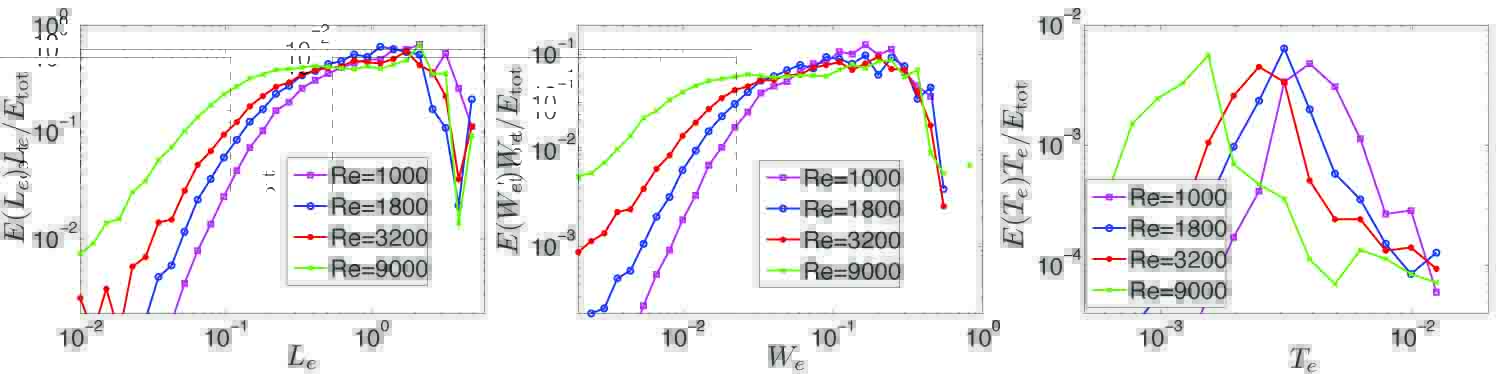}
 \caption{The compensated energy dissipation rate $E(X)X$ for Euclidean scales $X \in \{L_e,W_e,T_e\}$ (normalized to total energy dissipation rate $E_\text{tot}$), for $Re = 1000$ (magenta), $Re = 1800$ (blue), $Re = 3200$ (red), and $Re = 9000$ (green). The threshold is chosen so that $j_\text{thr}/j_\text{rms} = 3.75$, which gives a similar combined energy dissipation rate and volume occupied for structures independently of $Re$ (see Fig.~\ref{fig1}). The scales are measured in the units of $2\pi$. \label{fig4}}
 \end{figure*}

\begin{figure*}[t!]
 \includegraphics[width=2\columnwidth]{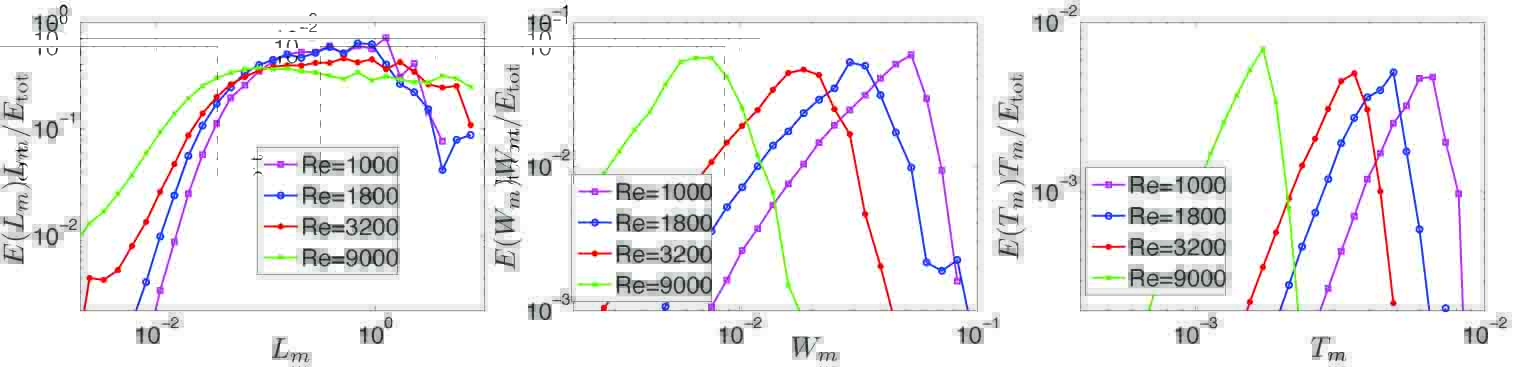}
 \caption{The compensated energy dissipation rate $E(X)X$ for Minkowski scales $X \in \{L_m,W_m,T_m\}$ (normalized to total energy dissipation rate $E_\text{tot}$), for $Re = 1000$ (magenta), $Re = 1800$ (blue), $Re = 3200$ (red), and $Re = 9000$ (green). Comparing with the Euclidean scales in Fig.~\ref{fig4}, the length and thickness scales in the two methods agree, but the intermediate scales exhibit different behavior. \label{fig4b}}
 \end{figure*}

For a more detailed study, we now directly determine the spatial scales at which intermittent energy dissipation takes place. Let $E(X) dX$ denote the combined energy dissipation rate for structures with scales between $X$ and $X + dX$, where $X \in \{L_e, W_e, T_e, L_m, W_m, T_m\}$ represents any of the characteristic scales. Then the maximum of the compensated energy dissipation rate, $E(X)X$, indicates at which $X$ most of the energy dissipation occurs. If we assume that $P({\cal E}) \sim {\cal E}^{-\alpha}$ and that ${\cal E} \sim X^\beta$ for arbitrary $\alpha$ and $\beta$, then energy dissipation will be distributed uniformly across all $X$ if and only if $\alpha = 2$. This follows from
\begin{eqnarray}
E(X) X \sim {\cal E}(X) P(X) X &\sim&  {\cal E}(X) \frac{d {\cal E}}{d X} P({\cal E}) X \nonumber \\
&\sim& X^\beta X^{\beta - 1} (X^{\beta})^{-\alpha} X \nonumber \\
&\sim& X^{\beta (2 - \alpha)} \,.
\end{eqnarray}

We first discuss the energy dissipated in the Euclidean scales. Shown in Fig.~\ref{fig4} are $E(L_e)L_e$, $E(W_e)W_e$ and $E(T_e)T_e$. Remarkably, energy dissipation is spread nearly uniformly amongst structures with $L_e$ and $W_e$ spanning intermediate to large scales. For $W_e$, this regime corresponds to inertial-range scales associated with the perpendicular energy cascade; for $L_e$, the scales are amplified by the anisotropy of the system (i.e. the ratio $B_0/b_\text{rms}$). There may be a small tendency for the energy dissipation to peak in structures with the largest scales, comparable to the system size; however, this tendency appears to decrease in the highest Reynolds number cases. The energy dissipated in these large scales does not change significantly with increasing~$Re$, although additional small scales are accessed due to a longer inertial range. In contrast to this, the energy dissipation is peaked at very small $T_e$ deep within the dissipation range, which accounts for energy dissipation at the bottom of the energy cascade. Energy dissipation peaks at smaller $T_e$ as~$Re$ is increased.

We now compare this to the energy dissipated in the Minkowski scales. Shown in Fig.~\ref{fig4b} are $E(L_m)L_m$, $E(W_m)W_m$ and $E(T_m)T_m$. As with the Euclidean case, energy dissipation occurs mainly in structures with $L_m$ spread throughout the inertial range and $T_m$ sharply peaked at small scales. However, unlike the Euclidean case where $W_e$ takes a continuum of inertial-range values, $W_m$ is strongly peaked at a scale between the intertial range and dissipation range. In fact, it appears that $W_m$ is representative of the dissipation scale.

The pronounced qualitative difference between $W_e$ and $W_m$ suggests that the two methods are measuring a different physical quantity. The Euclidean width by definition must be no greater than the perpendicular scale at broadest part of the structure. The fact that it lies in the inertial-range is then strongly indicative of the structure as a whole spanning inertial-range scales in the perpendicular direction. On the other hand, it is rather ambiguous what the Minkowski width could represent. A simple possibility is that the structure has an extended dissipation-scale tail which is measured by $W_m$. Alternatively, it is possible that $W_m$ is sensitive to dissipation-scale fluctuations along the structure, representing a characteristic scale for ripples or irregularities. In any case, it is not surprising that the dynamics responsible for the complex morphology of structures may favor the dissipation scale, since the energy cascade for current density peaks at the top of the dissipation range.

 \begin{figure}[t!]
\includegraphics[width=\columnwidth]{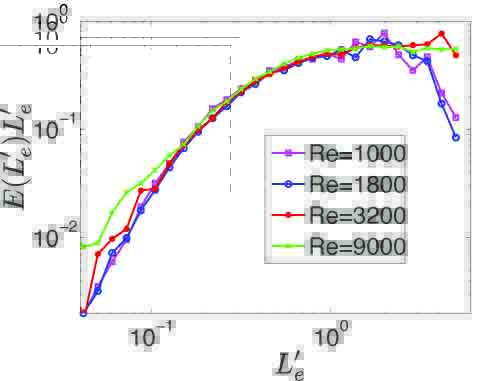}
 \includegraphics[width=\columnwidth]{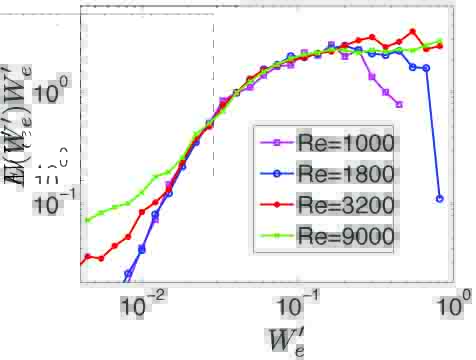}
 \caption{Energy dissipated at the rescaled length, $L_e' = L_e (Re/Re_0)^{0.65}$ and rescaled width $W_e' = W_e (Re/Re_0)^{0.85}$ (with arbitrary normalization). \label{figex}}
 \end{figure}

The energy distributions in Fig.~\ref{fig4} and Fig.~\ref{fig4b} exhibit unambiguous scaling behavior with Reynolds number, with all characteristic scales decreasing with $Re$. However, it is difficult to get a definitive quantitative measurement for these scalings due to the limited range in $Re$ and uncertainty into how to best normalize the distributions for proper comparison. For both methods, the lower cutoff for inertial-range lengths goes roughly as $L_\text{cutoff} \sim Re^{-\lambda}$ with $\lambda \approx 0.65 \pm 0.15$. This is demonstrated in Fig.~\ref{figex}, which shows the energy dissipated at rescaled Euclidean length, $L_e' = L_e (Re/Re_0)^{0.65}$ where $Re_0 = 1000$ is a reference scaling factor. The cutoff for inertial-range $W_e$ appears to have a somewhat different scaling, as $W_\text{cutoff} \sim Re^{-\omega}$ with $\omega \approx 0.85 \pm 0.10$, also shown in Fig.~\ref{figex}. Incidentally, the scaling for the peak of energy dissipation in $W_m$ is similar to this. The peak for thickness appears to scale with $Re$ in a similar way as the length cutoff; however, comparison with $512^3$ simulations suggest that the thickness measurements may be affected by resolution. If one interprets the length cutoff and width cutoff as dissipation scales in the parallel and perpendicular directions, respectively, then their scaling is consistent with critical balance, $L \sim W^{2/3}$ \citep{goldreich_sridhar1995, boldyrev2006}. However, a more complete theory is required to fully explain the observations.

\begin{figure}[t!]
 \includegraphics[width=\columnwidth]{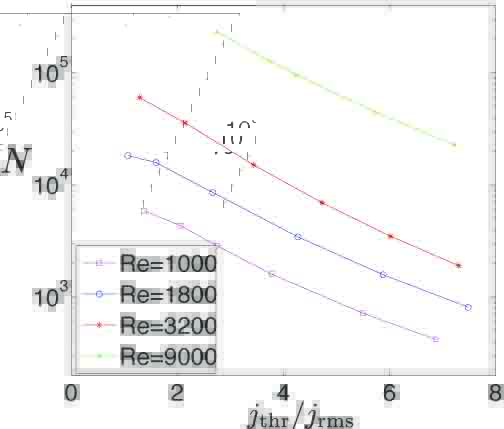}
 \caption{The filtered number of structures $N$ per snapshot as a function of the rescaled threshold for $Re = 1000$ (magenta), $Re = 1800$ (blue), $Re = 3200$ (red), and $Re = 9000$ (green). The number of structures at any given threshold increases strongly with $Re$. \label{fig5}}
 \end{figure}

Finally, we remark on the number of structures per snapshot. The simplest approach is to directly count the unfiltered number of structures in the population, $N$. However, this result is strongly skewed toward under-resolved structures near the threshold, which strongly contribute to $N$ even though they represent a negligible contribution to the total energy dissipation. To obtain a more reasonable estimate of the population size, we count only the structures with energy dissipation rates greater than a minimum value $C h^3 \eta j_\text{thr}^2$, where $h^3$ is the lattice volume element and $C \ge 1$ is some fixed number. This criterion removes many of the under-resolved, unphysical structures. We find that $N$ strongly increases with $Re$, as shown in Fig.~\ref{fig5} for $C = 8$. This trend is similar for other values of $C$ (including the unfiltered case of $C = 1$), and also for other filtering methods, e.g., volumetric filtering of structures or Fourier space filtering of the fields.

\begin{figure}[t!]
 \includegraphics[width=\columnwidth]{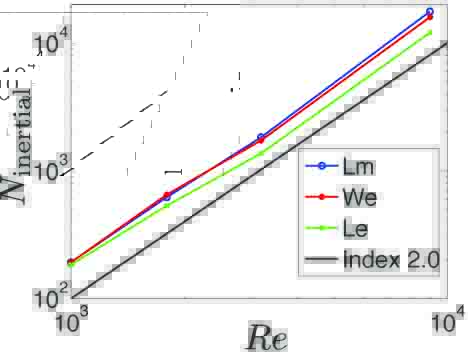}
 \caption{Number of inertial-range structures versus $Re$ for the given threshold of $j_\text{thr}/j_\text{rms} \approx 3.75$. We find that $N_\text{inertial} \sim Re^2$ using inertial-range populations for three different quantities: $L_e$ (green), $L_m$ (blue), and $W_e$ (red). \label{fig5b}}
 \end{figure}

Potentially more meaningful than the total population is the number of inertial-range structures, $N_\text{inertial}$. To determine this quantity, we count only structures in the flat region of the energy distributions in Fig.~\ref{fig4} and Fig.~\ref{fig4b}, i.e. with $L_e > L_\text{cutoff}$, $L_m > L_\text{cutoff}$, and $W_e > W_\text{cutoff}$, where $L_\text{cutoff}$ and $W_\text{cutoff}$ are the Reynolds-number dependent lower cutoff for the inertial-range in $E(L_e)L_e$ and $E(W_e)W_e$, with scalings as implied by Fig.~\ref{figex}. As shown in Fig.~\ref{fig5b} for the given threshold, we find that $N_\text{inertial} \sim Re^2$ for the inertial-range populations in all three distributions.

\section{Conclusions}

Due to enormous Reynolds and magnetic Reynolds numbers, direct numerical simulations of astrophysical turbulence are impossible. Therefore, analytical and numerical studies of the scaling of the physical quantities with the Reynolds number become extremely valuable \citep[][]{longcope_sudan1994,ng_etal2012,lin2013}.  In this work, we found the scaling associated with very intense dissipative structures in MHD turbulence, which convert into heat about 40\% of magnetic energy in about 2\% of the volume. We conclude that as resistivity of the system is decreased (or equivalently, $Re$ increased), the following scenario occurs. The lengths and widths of structures continue to occupy a continuum of large scales spanning the inertial range and often comparable to the system size. The thickness of structures decreases while the number of structures increases. Energy dissipation then takes place in a large number of thin, broad, tightly-packed current sheets. This suggests that the dissipative structures in this system may be classified both as nanoflares \it and \rm coherent structures. If we further extrapolate our results, the progressively increasing concentration of structures suggests a rather nontrivial limit of resistive MHD turbulence at infinitely large Reynolds and magnetic Reynolds numbers.

Our results suggest that energy dissipation rates may be distributed with the critical power law of index $-2.0$, so that the populations of weak structures and intense structures both contribute equally to the overall energy dissipation. Assuming that this distribution converges with higher Reynolds numbers, this lack of characteristic event type could potentially be exploited in future theoretical studies.

In addition to the above analysis of structures in the current density, we have applied our procedure on structures in the vorticity $\omega$, along with the associated viscous energy dissipation rate ${\cal E} = \int dV \nu \omega^2$. The vorticity structures in our simulations are found to be sheet-like, with similar statistical properties as the current sheets. The total viscous energy dissipation is comparable to but somewhat less than the resistive energy dissipation, consistent with the existence of residual energy \citep{wang11,boldyrev_pw2012}.

The methods presented in this work can be applied to MHD simulations with more specialized boundary conditions and forcing mechanisms, including the line-tied model for the solar corona and sheared-box model for accretion disks. Indeed, line-tied boundary conditions are thought to strongly affect current sheet formation \citep{ng_bhattacharjee1998, cowley_etal1997, zweibel_li1987} and magnetic tearing modes \citep{huang_zweibel2009,delzanno_finn2008}, particularly at global scales. It is therefore of interest to determine to what extent our present findings can be extrapolated to large scales and realistic parameters in those cases. These methods will also be applied to simulations of the kinematic and dynamic dynamos in order to determine the morphological differences between structures in the two cases.

\acknowledgements
This work was supported by the US DOE award DE-SC0003888, the DOE grant DE-SC0001794, the NSF grant PHY-0903872, the NSF Center for Magnetic Self-Organization in Laboratory and Astrophysical Plasmas at U. Wisconsin-Madison, and the Science and Technology Facilities Council (STFC) UK.  High Performance Computing resources were provided by the Texas Advanced Computing Center (TACC) at the University of Texas at Austin under the NSF-Teragrid Project TG-PHY080013N.

%\bibliographystyle{apj}
%\bibliography{refs_str}
%\bibliographystyle{plain}

%\bibliography{refs_str}

\end{document}